\documentclass[prd,twocolumn,nofootinbib,amsfonts]{revtex4-2}
\usepackage[utf8]{inputenc}
\usepackage{amssymb}
\usepackage{amsmath}
 \usepackage{natbib}
\usepackage{graphicx}
\usepackage{dcolumn}
\usepackage{bm}
\usepackage{leftidx}
\usepackage{txfonts}
\usepackage{color}
\usepackage{enumitem}
\usepackage{floatrow}
\usepackage{amsmath}
\RequirePackage{mathptmx}      
\RequirePackage{latexsym}
\DeclareUnicodeCharacter{2212}{-}

\newcommand{\old}[1]{}

\def\be{\begin{equation}}
	\def\ee{\end{equation}}
\def\bea{\begin{eqnarray}}
	\def\eea{\end{eqnarray}}
\begin{document}

\title{The impact of compact object deformation on thin accretion disk properties}


\author{Shokoufe Faraji}

\email{shokoufe.faraji@zarm.uni-bremen.de}


\affiliation{Waterloo Centre for Astrophysics, University of Waterloo, Waterloo, ON N2L 3G1, Canada}

 \affiliation{Department of Physics and Astronomy, University of Waterloo, Waterloo, ON N2L 3G1, Canada}

\affiliation{Perimeter Institute for Theoretical Physics, 31 Caroline Street North, Waterloo, N2L 2Y5, Canada}

\affiliation{Center of Applied Space Technology and Microgravity (ZARM), University of Bremen, 28359 Germany}


\begin{abstract}
We investigate the standard relativistic geometrically thin and optically thick accretion disk in the background of a deformed compact object. The main purpose of this work is to determine whether such a deformed object possesses its own observational fingerprint that can distinguish it from Schwarzschild and Kerr black holes. Our analysis reveals the properties of this relativistic accretion disk model and its dependence on the initial parameters.
\end{abstract}

\maketitle


\section{Introduction}



The study of accretion disks is fundamental to our understanding many high-energy astrophysical processes. These disks naturally are formed due to the conservation of angular momentum in material falling toward a strong gravitational center. Accretion disks are observed in various astrophysical systems, including around young stellar objects, main-sequence stars, and supermassive black holes, where they play a critical role in phenomena ranging from star formation to the high-energy emissions of active galactic nuclei (AGN) \cite{1969Natur.223..690L,1995cvs..book.....W,2011ARA&A..49..195A,2013A&ARv..21...69R}. Over the years, significant progress has been made in understanding accretion disks through both analytical models and numerical simulations in a complementary fashion. The standard thin accretion disk model initially introduced by Shakura \& Sunyaev \cite{1973A&A....24..337S} and has been further extended including Novikov \& Thorne \cite{1973blho.conf..343N}, Lynden-Bell \& Pringle  \cite{1974MNRAS.168..603L}, {Paczy{\'n}ski} \& Bisnovatyi-Kogan \cite{1981AcA....31..283P}, Muchotrzeb \& {Paczy{\'n}ski} \cite{1982AcA....32....1M}. 


In this work, we explore the structure of thin accretion disks within the context of spacetime including a quadrupole moment. From a dynamical point of view, the quadrupole parameter can be interpreted as a perturbation parameter  of the Schwarzschild spacetime.The spacetime we consider, known as the $\rm q$-metric is a simple yet powerful generalization of the asymptotically flat solutions to Einstein's equations that includes a quadrupole moment. The first static and axially symmetric solution with an arbitrary quadrupole moment was described by Weyl \cite{doi:10.1002/andp.19173591804}. Subsequently, Erez and Rosen \cite{osti_4201189} introduced a static solution with an arbitrary quadrupole in prolate spheroidal coordinates. Later, Zipoy \cite{doi:10.1063/1.1705005} and Voorhees \cite{PhysRevD.2.2119} discovered an equivalent transformation, simplifying the solution and allowing it to be treated analytically. Applying another transformation to this metric results in what is now known as the $\rm q$-metric, which we discuss briefly in Section \ref{sec:qmetric}.

In addition to the $\rm q$-metric, other solutions such as the Hartle-Thorne \cite{1968ApJ...153..807H}, Manko-Novikov \cite{1992CQGra...9.2477M}, Johannsen-Psaltis \cite{PhysRevD.83.124015}, and $\rm q$-Kerr \cite{2006CQGra..23.4167G} metrics account for various forms of deformation, including rotational effects and higher-order multipole moments. These metrics allow for both oblate and prolate configurations, making them useful for understanding how deviations from spherical symmetry impact accretion disk properties. In this work, we focus on the static, quadrupole $\rm q$-metric and investigate the properties of a thin accretion disk model in the presence of quadrupole moments and examine how these disks differ around a deformed compact object from those around static black holes. Consequently, the properties of an accretion disk surrounding this spacetime explicitly depend on the value of the quadrupole parameter. This dependency enables a clear distinction between the accretion disk properties a static or Kerr black hole and those of an oblate or prolate compact object. Previous works by the author and collaborators have explored the thick disk model in the background of the static $\rm q$-metric \cite{2021A&A...654A.100F}, as well as its rotating counterpart \cite{2022EPJC...82.1149F}.


This paper is organized as follows. In Section \ref{sec:qmetric} we explain the spacetime. In Section \ref{sec:disk} we revisit and discuss the relativistic thin disk model, and the construction of the thin disk within this metric is explained. The solutions and results are presented in Section \ref{sec:results}. Finally, the summary and conclusions are provided in Section \ref{sec:summary}. Throughout this paper, the geometrized units, $c=1$, $M=1$ and $G=1$ have been used, except for the results.

\section{Space-time}\label{sec:qmetric}
The $\rm q$-metric describes static, axially symmetric, and asymptotically flat solutions to the Einstein field equations. By incorporating the quadrupole moment, the $\rm q$-metric accounts for the deformation of the gravitational field due to mass distribution, which is not captured by the spherically symmetric Schwarzschild solution.  Consequently, the $\rm q$-metric characterizes the exterior gravitational field of an isolated static axisymmetric mass distribution, providing a more accurate description of astrophysical objects with non-spherical mass distributions. 

In fact, the presence of a quadrupole can change the geometric properties of space-time (see e.g. \cite{2021A&A...654A.100F}). The quadrupole moment introduces additional terms in the metric that describe the deviation from spherical symmetry. These terms also affect the timelike and lightlike trajectories. As a result, the $\rm q$-metric is particularly useful for modeling the gravitational fields of realistic astrophysical bodies where the effects of higher-order multipole moments cannot be ignored. The metric in Schwarzschild-like coordinates $(t, x, y, \phi)$ is presented as follows \cite{osti_4201189,PhysRevD.39.2904}



\begin{align}
   ds^2 = &\left(1-\frac{2M}{r}\right)^{1+{\rm q}} dt^2 - \left(1-\frac{2M}{r}\right)^
    {-{\rm q}} \nonumber\\
    &\left[ \left(1+\frac{M^2\sin^2\theta}{r^2-2Mr}\right)^{-{\rm q}(2+{\rm q})}\left(\frac{dr^2}{1-\frac{2M}{r}}+r^2d\theta^2\right)\right. \nonumber\\
  &\left.+r^2\sin^2\theta d\phi^2\right].\
\end{align}
where $t \in (-\infty, +\infty)$, $r \in (2M, +\infty)$, $\theta \in [0,\pi]$, and $\phi \in [0, 2\pi)$, also $M$ is a parameter that can be identified as the mass of the body generating the field, which is expressed in the dimension of length. This metric contains two free parameters, namely the total mass, and quadrupole moments ${\rm q}$, which are taken to be relatively small. In the case of ${\rm q}=0$, the Schwarzschild metric is recovered. Furthermore, values of ${\rm q} > 0$ describe the exterior field of an oblate central object, while ${\rm q} < 0$ is related to a prolate source.

In fact, the $\rm q$-metric, has a central curvature singularity at $r=0$. Furthermore, another singularity is situated at a finite distance from the origin at $r=2M$ for any chosen value of quadrupole. However, considering the relatively small quadrupole moment ${\rm q}$, this singularity is located very close to the origin and can be covered by a proper interior solution in such a way that out of this region the metric is asymptotically flat \cite{Quevedo:2010vx}. Therefore, this solution can describe the exterior gravitational field of a deformed compact object up to the quadrupole. By Geroch definition \citep{1970JMP....11.2580G} for multipole moment and for avoiding a negative mass distribution, the quadrupole is restricted at most to the domain of $\rm {\rm q}\in(-1,\infty)$ \footnote{It is worth mentioning that the difference in the classical and relativistic multipole moments appears at first in the octupole moment. However, we can always choose an origin in such a way that dipole vanishes, then the next term that shows deviation from the classical one will be revealed in the sixteen-pole moment.}. 

Furthermore, in the study of circular geodesics${\rm q}$ in the equatorial plane, we have real values for angular momentum for ${\rm q} \in [-1+\frac{\sqrt{5}}{5}, \infty)$, and we interpret as the physical region in the scope of general relativity. Furthermore, studying astrophysical systems allows for tighter constraints on the physically suitable values of the quadrupole parameter. The place of ISCO in Schwarzschild is at $r=6M$. For a negative ${\rm q}$ corresponding to the slightly prolate object, ISCO is closer to the horizon $r=2M$, and for a positive ${\rm q}$ related to the slightly oblate object, the place of ISCO is pushed away. This behavior contrasts with the Hartle-Thorne metric, where rotational effects lead to different ISCO shifts. Similarly, the Manko-Novikov metric, which includes arbitrary multipole moments, can exhibit more complex ISCO behavior due to the presence of higher-order moments. In what follows, we review the construction of the thin accretion disk.



\section{Relativistic thin accretion disk }\label{sec:disk}

The standard thin disk model has been used to explain a variety of observations where the gas is cold and neutral, making the coupling between the magnetic field and the gas negligible (e.g., \cite{1994ApJ...421..163B}). This model has proven effective in scenarios where magnetic influences are minimal, allowing for simpler analyses of accretion processes. However, when considering the effects of quadrupole moments, it becomes essential to revisit these models and account for potential deviations in the gravitational field structure. By doing so, we can enhance our understanding of the dynamics and properties of accretion disks in more complex gravitational potential.

\subsection{Assumption of thin disk models}

In the standard thin accretion disk model, we assume a steady, axisymmetric fluid configuration. The model further suggests that the disk can locally radiate a significant portion of its rest mass energy as thermal black body-like radiation, generated through a viscosity mechanism. As a result, the thin disk is considered a cold accretion disk compared to the virial temperature, with gas temperatures around 100 K, depending on the mass of the central object.

One of the key assumptions of the standard thin disk model is that it is extremely thin and confined to the equatorial plane, implying that the ratio of the disk half-thickness \(H = H(r)\) to the radius \(r\) is very small. As a result of the geometrically thin assumption, effectively, all physical quantities depend on the vertical distance from the equatorial plane and the central object's radial distance. Thus, the two-dimensional disk structure can be decoupled to two one dimensional configurations, namely a radial quasi-Keplerian flow and a vertical hydrostatic structure, and one can drop the $z$ dependence of the model by considering vertically integrated quantities, as we will see in the next subsection. Additionally, the generated heat and radiation losses are in balanced $\rm Q^{gen} = Q^{rad}$, and caused to have a negligible advection, since ${\rm Q}^{\rm adv}\sim(H/r)^2{\rm Q}^{\rm gen}$.  
Consequently, it causes to luminosities be approximately below $30\%$ of the Eddington luminosity, $L_{\rm Edd}= 1.26\times 10^{38}\left(\frac{M}{M_{\odot}}\right)$. Above this limit, the gas becomes optically too thick and can not radiate all the dissipated energy locally \citep{2006ApJ...652..518M,2008ApJ...676..549S}. 
Therefore, one should justify applying the standard thin model to disks with higher luminosity.

Furthermore, the model assumes quasi-Keplerian circular orbits with a small radial drift velocity $u^r$ much smaller than the angular velocity. Near the central compact object $u^r$ becomes negative, resulting in mass accretion. The inner edge of a thin accretion disk with sub-Eddington luminosities corresponds to the region from which most of the luminosity originates. It is widely agreed that, with very high accuracy, this inner edge is located at the Innermost Stable Circular Orbit (ISCO). This model introduces viscosity through a so-called ${\alpha}$-prescription. It states that shear stress is supposed to be a form of viscosity responsible for transporting angular momentum and energy outward, accreting matter inward, also heating the gas locally. It is given by $S_{\hat{r}\hat{\phi}}={\rm q} P$, where $S_{\hat{r}\hat{\phi}}$ is the only non-vanishing component of viscous (internal) stress-energy tensor in the fluid frame, and $P$ is pressure \footnote{There is also another version of this prescription that is commonly used in terms of the vertically averaged sound speed and the vertical half-thickness of the disk, as $\nu\simeq{\rm q} c_s H$; however, one should take it by cautious \citep{blaes2002physics}.}. Almost in accretion disk models, it is assumed that the dimensionless parameter ${\rm q}$ is a constant in this range $[0.01-0.1]$, e.g., \cite{RevModPhys.70.1} that has a very good agreement with simulations.  In the following part, we briefly state the equations governing the dynamics of the thin disk model.



%

\subsection{Equations of thin disk models}\label{sec:equation}

Three fundamental equations govern the radial structure of the thin disk model. First, the particle number conservation
\begin{equation}\label{restmasscon}
(\rho u^{\mu})_{;\mu}=0 \,,
\end{equation}
where $u^{\mu}$ is the four-velocity of the fluid and $\rho$ is the rest mass density. The mass accretion rate is connected to this conservation law, indicating that the accretion rate should be constant. Otherwise, we would observe matter accumulating in certain regions of the disk. The two other equations are related to the stress-energy conservation $T^{\mu \nu}{}_{;\nu}=0$. One is described by the radial component of conservation of energy-momentum tensor, parallel to the four-velocity 
\begin{equation}\label{energycon}
u_{\mu} T^{\mu \nu}{}_{;\nu}=0\,.
\end{equation}
Another is the radial component of projection of this conservation onto the surface normal to the four-velocity
\begin{equation}\label{NSE}
h_{\mu \sigma}(T^{\sigma \nu})_{;\nu}=0 \,,
\end{equation}
where $h^{\mu \nu} = u^{\mu} u^{\nu} + g^{\mu \nu}$ is the projection tensor giving the induced metric normal to $u_{\mu}$. The stress-energy tensor $T^{\sigma \nu}$ reads as
\begin{align}
T^{\mu\nu}=hu^{\mu}u^{\nu}-pg^{\mu\nu}+q^{\mu} u^{\nu}+q^{\nu} u^{\mu}+\tau^{\mu\nu},
\end{align}
where $h$ is enthalpy density, which is the sum of internal energy per unit proper volume and pressure over rest-mass density, $p$ is the pressure, $q^{\nu}$ is transverse energy flux, and $\tau^{\mu\nu}$ is the viscous stress-energy tensor. In relativistic form, it is proportional to the shear rate, where the only non-vanishing component, according to assumptions of the thin disk model, is $\tau^{r\phi}=\alpha P$ and measures the rate of change of the angular velocity with the radius. 

By applying the assumptions of the thin disk model to the basic equations \eqref{restmasscon}-\eqref{NSE}, along with the relations describing radiative energy transport and the vertical pressure gradient, we derive a system of nonlinear algebraic equations governing this model \citep{1973blho.conf..343N}, as follows. For steady accretion through the thin disks, from the continuity equation, the mass accretion rate is given by
\begin{align}
\dot{M}=-2\pi r u^r \Sigma=\text{constant}.    
\end{align}
So, the radial velocity of the fluid then reads as

\begin{align}\label{massrate}
u^r=-\frac{\dot{M}}{2\pi r \Sigma}.
\end{align}
Although the motion is nearly circular in the disk, bypassing the ISCO, the radial velocity increases rapidly. The surface density $\Sigma$ is obtained by vertical integration of the density
\begin{align}\label{sigma2}
\Sigma=\int^{+H}_{-H}\rho {\rm d}z =  2\rho H,
\end{align}

where $H$ is disk height or half of the thickness of the disk. Based on the assumptions of thin disk models, heat flow is considered to be primarily in the vertical direction. Consequently, the time-averaged flux of radiant energy (energy per unit proper area and proper time) emitted from the upper and lower surfaces, denoted as $F$, is related to the vertical heat flow \cite{1973blho.conf..343N,1974ApJ...191..499P} as
$q^z(r,z) = F(r) \frac{z}{H(r)}$. Applying the assumptions to the fundamental equations \eqref{restmasscon}, \eqref{energycon}, and \eqref{NSE} allows us to readily obtain

\begin{align}\label{ene}
\frac{(\Omega {L}-E)^2}{\Omega_{,{r}}}\frac{{F \sqrt{-|g|}}}{{\dot{M}}}= \int_{{r}_0}^{r} \frac{(\Omega {L}-{E})}{4\pi}{L}_{,{r}} d{r},  
\end{align}
where $E=-u_t$ and $L=u_\phi$ are the specific energy and angular momentum in the equatorial plane, and $\Omega=\frac{u^{\phi}}{u^t}$ is the corresponding angular velocity. The vertically integrated viscous stress $W$, using ${\alpha}$-prescription is given by 
\begin{align}\label{w}
W=\int^{+H}_{-H}\tau{}^{\phi r} {\rm d}z = 2{\alpha}P {H}.
\end{align}
And the energy flux reads as
\begin{align}\label{navi}
F = -\int^{+H}_{-H} \sigma_{\phi r}\tau{}^{\phi r} {\rm d}z = -\sigma_{\phi r}{W},
\end{align}
where $\sigma_{\phi r}$ is the off-diagonal part of the shear tensor in the fluid frame, given by
\begin{equation}\label{viscosity}
\sigma_{\phi r} = \frac{1}{2}(u_{\alpha ; \mu} h_{\beta}^{\mu} + u_{\beta ; \mu} h^{\mu}_{\alpha}) -\frac{1}{3}h{}_{\phi r}u^{\beta}{}_{;\beta}.
\end{equation}
According to the assumptions, at each radius, the emission is like black-body radiation. Therefore, the energy transportation is given by
\begin{equation}\label{OD}
a T^4={\Sigma} {F}{\kappa},
\end{equation}
here $\kappa$ is Rosseland-mean opacity, where Neglecting line opacity and bound-free opacity, here we consider

\begin{align}
\kappa&=0.40 \nonumber\\
&+ 0.64\times10^{23} \left(\frac{\rho}{{\rm g} \quad cm^{-3}}\right)\left(\frac{T}{K}\right)^{-\frac{7}{2}} cm^2{\rm g}^{-1}\nonumber,\
\end{align}
where the first term is electron scattering opacity, the second one is free-free absorption opacity, also $a$ is the radiation density constant given by $a= \frac{4 \sigma}{c}$, where $\sigma$ is Stefan-Boltzmann’s constant. The pressure $P$ is the sum of gas pressure from nuclei and the radiation pressure,
\begin{equation}\label{P}
P=\frac{\rho kT}{m_p}+\frac{a}{3}T^4,
\end{equation}
where ${m_p}$ is the rest mass of the proton, ${k}$ is Boltzmann's constant, $a$ is the radiation density constant, and $T$ is the temperature \footnote{In fact, we ignore the mass difference between neutrons and protons in the first term for simplicity.}. In practice, the pressure equation in the vertical direction is given by

\begin{equation}\label{VP}
\frac{P}{\rho}=\frac{1}{2}\frac{(HL)^2}{r^4}\,,
\end{equation}
which is derived from the relativistic Euler equation. By solving these equations \eqref{massrate}-\eqref{VP}, one can calculate physical quantities subsequently. Besides, there are there free parameters in the model describing a thin disk solution, namely $M$, mass accretion rate $\dot{M}$, and viscosity parameter ${\rm q}$. Note that, this system of equations admits a unique solution for considering positive temperature and positive pressure. 

\subsection{Different regions of Accretion disk}

The thin disk is constructed with three different regions, allowing for the derivation of three separate local solutions. These solutions can be distinguished based on whether gas pressure or radiation pressure is dominant and whether opacity is dominated via electron scattering or free-free absorption. However, the qualitative features of the global solution, which can be obtained by patching the local solutions, depend significantly on the inner region close to the ISCO. There are three possibilities.

\begin{enumerate}[label=\Roman*.]

    \item {Inner region of accretion disk} Radiation pressure-electron scattering dominated solution: $P \simeq P_{rad}$ and $\kappa \simeq  \kappa_{es}$.

    \item {Middle region of accretion disk} Gas pressure-electron scattering dominated: $P \simeq P_{gas}$ and $\kappa \simeq \kappa_{es}$.

    \item {Outer region of accretion disk} Gas pressure-free free absorption dominated solution: $P \simeq P_{gas}$ and $\kappa \simeq \kappa_{ff}$.

\end{enumerate}


\begin{figure}
    \includegraphics[width=1.05\hsize]{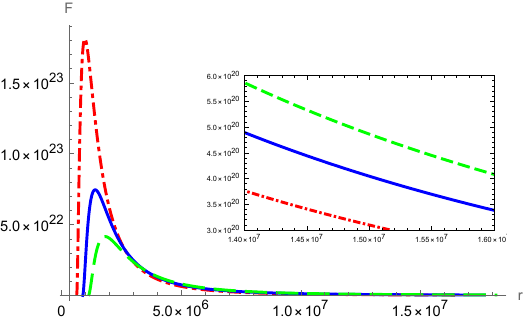}
    \includegraphics[width=1.05\hsize]{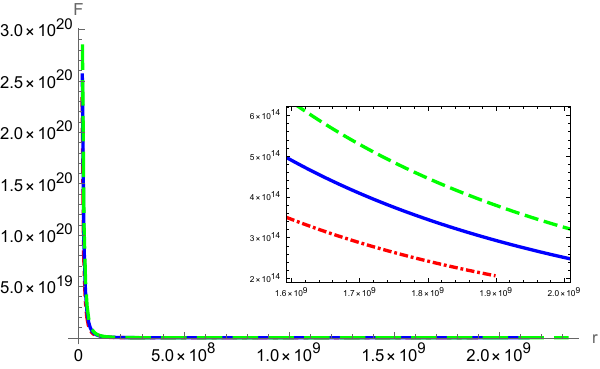}
    \includegraphics[width=1.05\hsize]{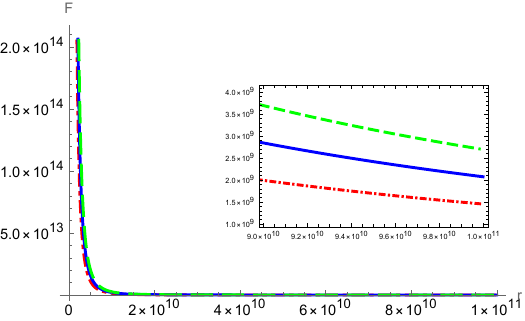}
     \caption{The flux profiles are shown for the inner, middle, and outer regions of the thin accretion disk, comparing different values of the quadrupole moment q. The green line represents $\rm q=0.3$, the blue line corresponds to Schwarzschild case $\rm q=0$, and the red line indicates $\rm q=-0.3$. Notably, the flux amplitude is higher for negative $\rm q$ values in the inner region close to the edge of the disk, while the differences between the curves decrease in the outer regions as the gravitational influence of the central object is reduced.}
    \label{fig:f-comparison3}
\end{figure}

\section{Results and discussion}\label{sec:results}
\begin{figure}
    \includegraphics[width=1.05\hsize]{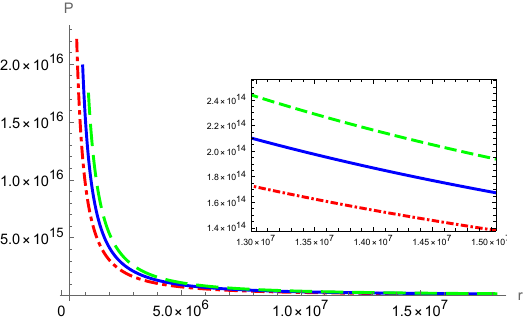}
    \includegraphics[width=1.05\hsize]{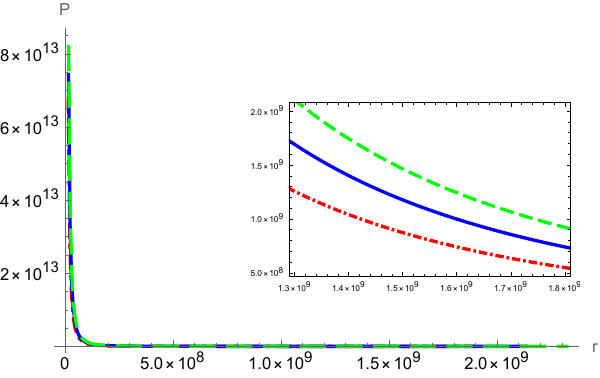}
    \includegraphics[width=1.05\hsize]{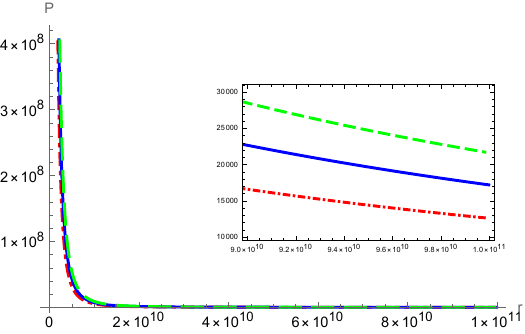}
     \caption{The pressure profiles for the inner, middle, and outer regions. The green line represents $\rm q=0.3$, the blue line shows Schwarzschild case $\rm q=0$, and the red line corresponds to $\rm q=-0.3$. The pressure is more intense for negative $\rm q$ values near the ISCO, while differences between the profiles diminish in the outer regions.}
    \label{fig:p-comparison3}
\end{figure}

\begin{figure}
    \includegraphics[width=1.04\hsize]{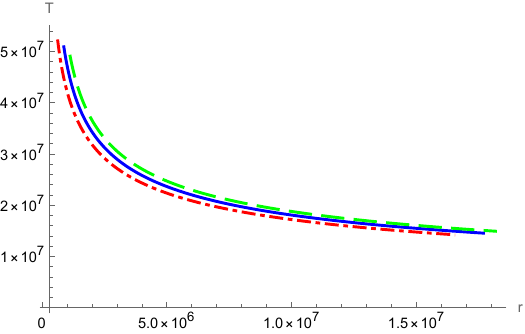}
    \includegraphics[width=1.04\hsize]{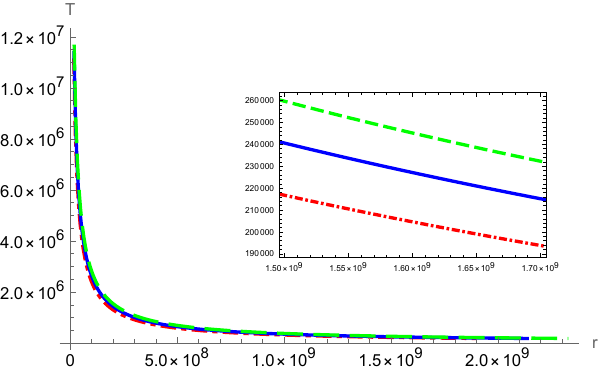}
    \includegraphics[width=1.05\hsize]{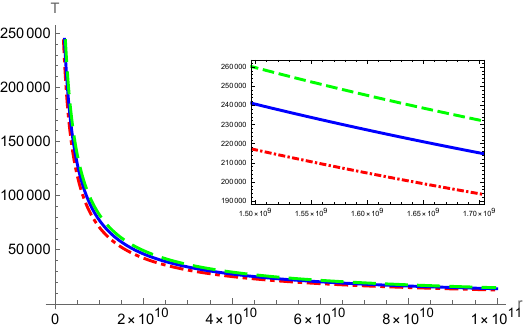}
     \caption{The temperature profile for the inner, middle and outer regions, respectively. The green line corresponds to $\rm q=0.3$, the blue line shows Schwarzschild case $\rm q=0$, and the red line represents $\rm q=-0.3$. The temperature decreases smoothly from the inner to the outer regions, reflecting the thermal equilibrium assumed in the disk.}
    \label{fig:t-comparison3}
\end{figure}

\begin{figure}
    \includegraphics[width=\hsize]{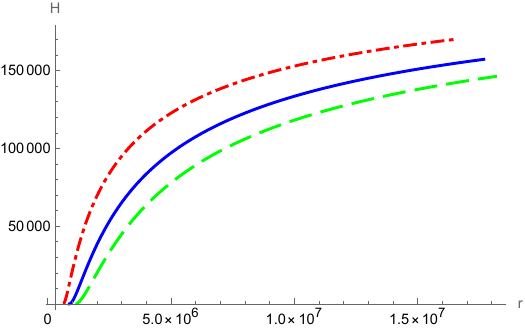}
    \includegraphics[width=\hsize]{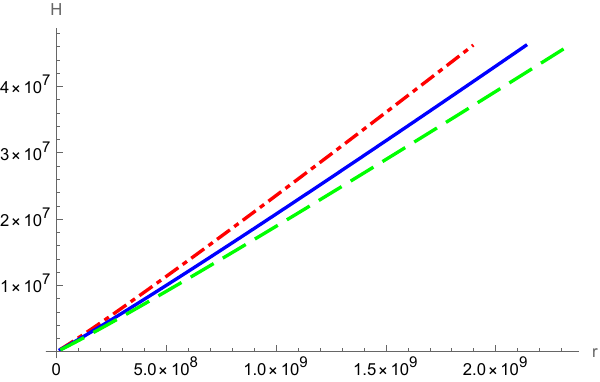}
    \includegraphics[width=\hsize]{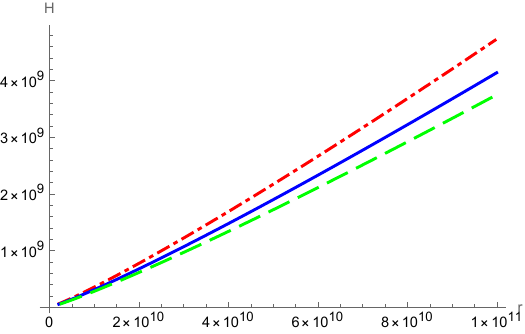}
     \caption{The disk height $H$ profile for the inner, middle and outer regions, respectively. The green line represents $\rm q=0.3$, the blue line shows Schwarzschild case $\rm q=0$, and the red line  represents negative values. The height increases with radial distance, with the disk becoming thicker outward, particularly for negative values of $\rm q$. }
    \label{fig:bh-comparison3}
\end{figure}

\begin{figure}
    \includegraphics[width=\hsize]{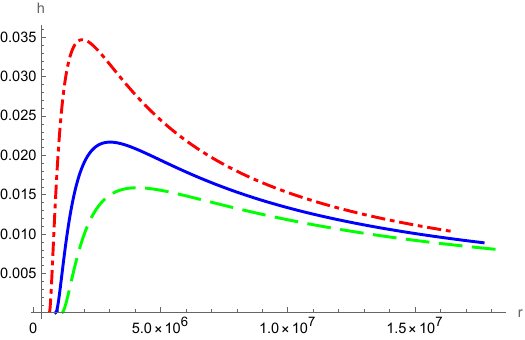}
    \includegraphics[width=\hsize]{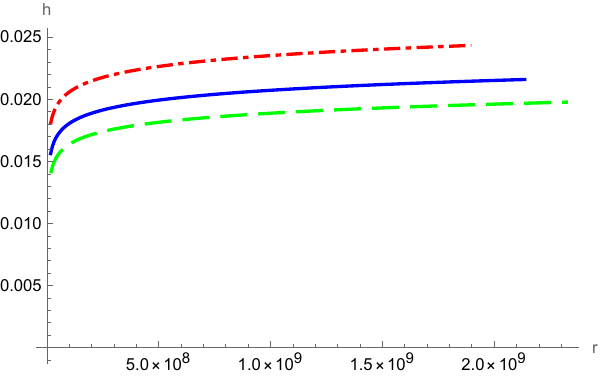}
    \includegraphics[width=\hsize]{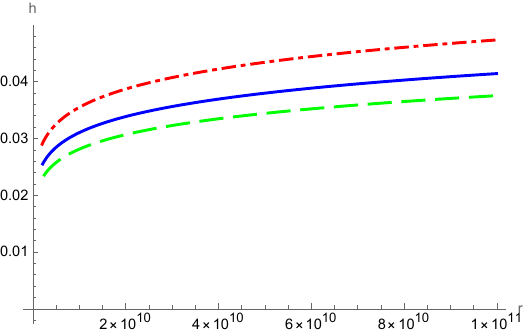}
     \caption{The disk geometrical thickness ratio $h$ profile profiles for the inner, middle, and outer regions. The green line $\rm q=0.3$, the blue line Schwarzschild case $\rm q=0$, and the red line $\rm q=-0.3$. The ratio illustrates how how the disk's relative thickness varies with radius, where negative $\rm q$ values result in a relatively thicker disk.}
    \label{fig:h-comparison3}
\end{figure}

\begin{figure}
    \includegraphics[width=1.03\hsize]{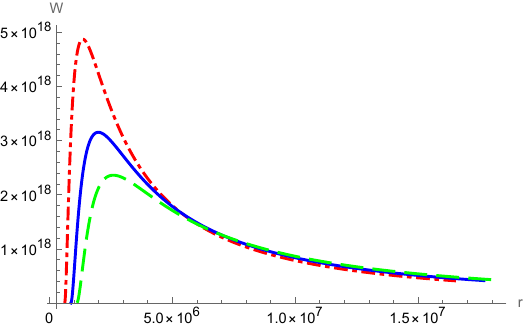}
    \includegraphics[width=1.03\hsize]{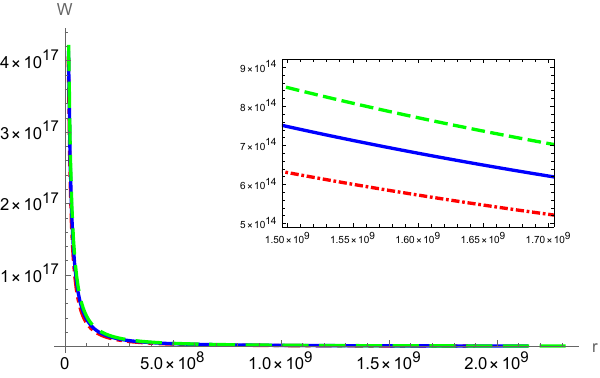}
    \includegraphics[width=1.03\hsize]{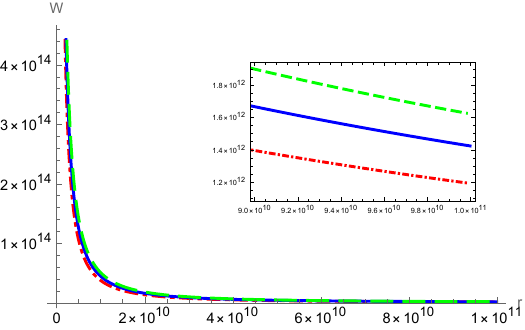}
     \caption{The viscous stress profile for the inner, middle and outer regions, respectively. The green line corresponds to $\rm q=0.3$, the blue line shows Schwarzschild case $\rm q=0$, and the red line represents $\rm q=-0.3$. The profiles show a strong similarity to the flux profiles due to their relationship through shear stress and angular momentum transport in the disk.}
    \label{fig:w-comparison3}
\end{figure}

\begin{figure}
    \includegraphics[width=1.06\hsize]{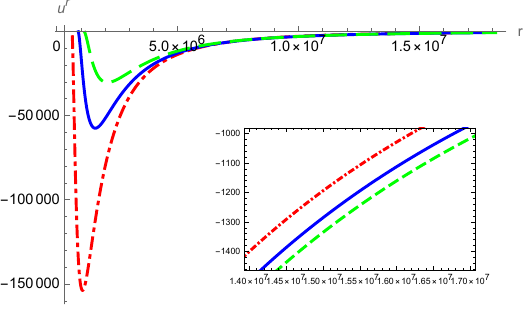}
    \includegraphics[width=1.06\hsize]{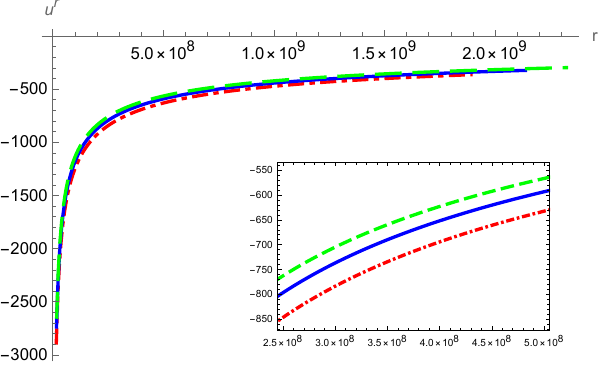}
    \includegraphics[width=1.06\hsize]{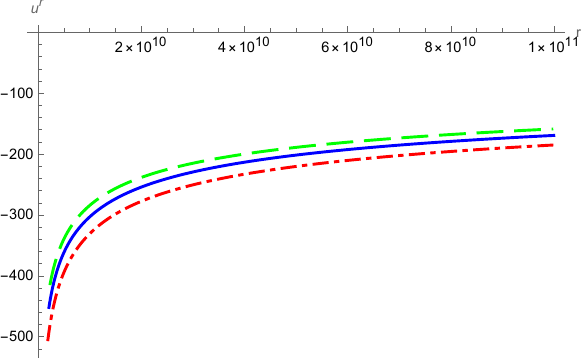}
     \caption{The radial drift velocity $u^r$ profile for the inner, middle and outer regions, respectively. The green line represents  $\rm q=0.3$, the blue line shows Schwarzschild case $\rm q=0$, and the red line corresponds to $\rm q=-0.3$. The drift velocity increases sharply near the ISCO due to stronger gravitational forces, leading to more rapid inward accretion.}
    \label{fig:ur-comparison3}
\end{figure}

The numerical solutions presented in the following figures were produced using the following physical constants and parameters

\begin{align}\label{param}
G&=6.67 \times 10^{-8} \quad {\rm cm^3 g^{-1} sec^{-2}},\\
c&= 3 \times10^{10} \quad {\rm cm \hspace{0.1cm} sec^{-1}}, \nonumber\\
k &=1.38 \times 10^{-16} {\rm erg \quad K^{-1}},\nonumber \\
a &= 7.56 \times 10^{-15} {\rm erg(cm^{-3}K^{-4})}, \nonumber\\
m_{p}&=1.672 \times 10^{-24} {\rm g},\nonumber\\
\kappa_{es}&= 0.40 \quad {\rm cm^2 g^{-1}},\nonumber\\
M&={M_{\odot}}(\sim 1.99 \times 10^{33} \rm g),\nonumber\\
{\alpha}&=0.01,\nonumber\\
\dot{M}&=1.33\times10^{17} {\rm erg\,s \,cm^{-2} },\nonumber\
\end{align}
with all results expressed in the CGS unit system. The results are highly sensitive to variations in parameters, particularly the $\alpha$ parameter and the mass accretion rate. It is crucial to note that for the thin accretion disk model to remain valid, the system must operate near the Eddington limit. In Figures \ref{fig:f-comparison3}-\ref{fig:ur-comparison3}, the main astrophysical quantities are plotted in there different regions of thin accretion disk. 

Figure \ref{fig:f-comparison3} presents the flux profiles for different values of the parameter ${\rm q}$. The Schwarzschild case is represented by the black line, while the flux amplitudes for prolate (${\rm q<0}$) and oblate (${\rm q>0}$) configurations are shown in red and green, respectively. In the inner region, the flux amplitude is significantly higher for negative ${\rm q}$ values, indicating a prolate source, compared to both the Schwarzschild case and the oblate configuration. This is primarily due to the location of the ISCO, which is considered the inner edge of the thin accretion disk. For negative values of ${\rm q}$, the ISCO is located closer to the central object. As the value of ${\rm q}$ increases, the ISCO moves outward. Consequently, for ${\rm q}<0$, the inner part of the disk is situated nearer to the central object, resulting in more intense characteristics across all quantities, including the flux. This inward shift for negative ${\rm q}$ values leads to higher energy release and, thus, a higher flux amplitude due to the stronger gravitational influence near the central mass. In the inner region, after the flux reaches its maximum, there is a change in the relative order of the profiles. For positive ${\rm q}$, the flux becomes more pronounced than for negative ${\rm q}$  as we move outward. However, these differences remain moderate because the farther regions are less influenced by the strong gravitational pull near the central object. This pattern of diminishing flux amplitude continues in the middle and outer regions, as the flux profiles gradually converge.

Figures \ref{fig:p-comparison3} and \ref{fig:t-comparison3} present the pressure and temperature profiles, respectively. Both profiles exhibit similar patterns, with more intense behavior near the ISCO for negative $\rm q$ values. As in the case of the flux, the positive quadrupole values dominate beyond the ISCO, although the differences between the profiles remain relatively small. The pressure profile, which is directly influenced by the density and velocity of the gas in the disk exhibits a sharper decrease in the inner regions. Conversely, the temperature profile exhibits a smoother decline from the inner region to the outer region. This can be attributed to the complex interplay between gravitational and thermal processes, more importantly, the thin disk model's assumption of thermal equilibrium, where the temperature at each radius is balance between heating due to viscous and radiative cooling, leading to a more gradual temperature variation.



Figure \ref{fig:bh-comparison3} shows the disk hight $H$. As expected, the height increases as one moves outward from the central object, indicating that the disk becomes geometrically thicker. The increase is more pronounced for negative $\rm q$ values (prolate case), where the disk is thicker at all radii compared to the Schwarzschild case and positive $\rm q$  values (oblate case). This behavior is linked to the stronger gravitational influence in the inner region for $\rm q<0$, which compresses the disk closer to the central object. Figure \ref{fig:h-comparison3} presents the ratio $h=H/r$, which represents the geometrical thickness of the disk relative to its radial extent. Unlike the disk height $H$, this ratio shows that the inner regions are relatively thinner compared to the outer regions where the disk becomes geometrically thicker. As with the height $H$ the relative thickness is higher for negative $\rm q$ values, while for positive $\rm q$ the disk remains thinner. This ratio also helps visualize how the disk’s thickness evolves in different parts of the accretion disk. It is important to note that the relative thickness should remain small, consistent with one of the main assumption of the standard thin disk model, which requires that the height be much smaller than the radial extent at all radii.


Figure \ref{fig:w-comparison3} shows the vertically integrated viscous stress, $W$. Its behavior closely resembles that of the flux profile. This similarity arises from their relationship, as expressed in equation \eqref{navi}, where they are connected through shear stress. In an accretion disk, shear stress represents the tangential force per unit area caused by differential rotation. The viscous stress, is related to the rate of angular momentum transport within the disk. Since the energy dissipation reflected in the flux is driven by this angular momentum transport, the flux and viscous stress profiles are inherently linked. Consequently, changes in the viscous stress directly affect the flux, leading to similar patterns in their profiles.


Figure \ref{fig:ur-comparison3} presents the behavior of radial drift velocity $u^r$, which plays a critical role in the accretion process. As the disk approaches the ISCO, the drift velocity typically becomes increases in magnitude (with negative values indicating inward motion toward the central object). The behavior of $u^r$, is strongly influenced by the quadrupole moment $\rm q$ especially in inner region, with negative $\rm q$ (prolate configurations) showing steeper velocity profiles, indicating more rapid inward motion. This is due to the stronger gravitational forces and steeper potential gradient near the ISCO in the presence of a prolate source. In contrast, for positive $\rm q$ (oblate configurations), the inward motion is more gradual, with a shallower velocity profile. These differences highlight how the quadrupole moment alters the dynamics of accreting material.

In general, as expected, we observe more pronounced behavior for negative quadrupole values compared to the Schwarzschild and positive quadrupole cases, due to the shift in the ISCO location, as previously discussed. As a result, accreting material experiences stronger gravitational forces, leading to sharper variations in physical quantities such as flux, pressure, and temperature in the inner regions of the disk. This proximity to the central object enhances the energy dissipation, resulting in a more luminous and dynamic accretion disk. Conversely, for positive quadrupole values, the ISCO shifts outward, causing the gravitational influence to weaken in the inner regions. This leads to more gradual changes in disk properties, with a less intense accretion flow and smoother variations in the corresponding physical quantities. These trends highlight the critical role that the quadrupole moment plays in shaping the structure and behavior of the accretion disk.

\section{Discussion and Conclusion}\label{sec:summary}

In this paper, we analyzed the structure of the thin accretion disk around a deformed compact object described by $\rm q$-metric. We have shown that the presence of a quadrupole moment can significantly changes the properties of the accretion disk leading to distinct spectral features. More precisely, the intensity of the physical quantities in this background is generally higher (lower) than the Schwarzschild solution for ${\rm q}<0$ (${\rm q}>0$) in the inner part region of the disk. In particular,the disk around a prolate source is more luminous at small radii compared to one around a regular black hole, with the effects diminishing monotonically as $\rm q$ increases near the ISCO.

In general, the quadrupole moments provides the possibility of taking them as an additional physical degree of freedom that can potentially link observational data to the theoretical models. Interestingly, when comparing this behavior to the Kerr solution where the rotation parameter $a$ replaces quadrupole parameter ${\rm q}$, we observe similar trends. For the co-rotating case ($a > 0$) the profiles are smaller than for the Schwarzschild case ($a = 0$) and the counter-rotating case ($a < 0$). However, the key difference is that in the Kerr background, this pattern holds across the entire disk, not just in the inner regions, without a reversal at larger radii. In contrast, the $\rm q$-metric shows more complex variations near the central object. Therefore, this distinction highlights notable differences in the behavior of profiles around a deformed compact object with a quadrupole moment and a rotating black hole.

It is worth noting that, the temperature profiles in the $\rm q$-metric show sharper gradients near the ISCO for $\rm q<0$, similar to the trends seen in the $\rm q$-Kerr and Johannsen-Psaltis metrics, where both quadrupole moments and rotational effects influence the disk's thermal structure. In contrast, the Manko-Novikov metric's more complex multipole structure may lead to additional features in the temperature profile, especially at small radii.

While this study demonstrates that the quadrupole moment in the 
$\rm q$-metric leads to distinct changes in accretion disk properties, particularly in the ISCO location, energy efficiency, and temperature distribution, it is important to acknowledge the limitations of the $\rm q$-metric. As mentioned, the 
$\rm q$-metric is constrained to the quadrupole order, limiting its applicability in systems where higher-order multipoles, such as octupole or hexadecapole moments, may become significant. In more complex systems, like neutron stars or rapidly rotating black holes, these higher-order moments could influence the disk structure, affecting the ISCO location, energy extraction efficiency, and radiative properties. Moreover, it represents a static, axially symmetric spacetime and thus neglects rotational effects. Since compact objects like black holes and neutron stars are typically rotating, the absence of angular momentum in the $\rm q$-metric means it cannot capture frame-dragging or the interaction between rotation and the quadrupole moment. This omission could lead to differences in observable quantities, such as flux and temperature profiles, compared to rotating spacetimes like the Kerr, Hartle-Thorne, or $\rm q$-Kerr metrics. Additionally, the $\rm q$-metric assumes an asymptotically flat spacetime, which may not reflect realistic astrophysical environments influenced by external fields, such as those from companion objects or accretion disks. While the $\rm q$-metric offers insights into quadrupole deformations, incorporating higher-order moments and rotation will be essential for modeling systems with more complex gravitational fields. Therefore, comparing these results with those predicted by the Hartle-Thorne, Manko-Novikov, Johannsen-Psaltis, and q-Kerr metrics highlights how quadrupole moments and rotational effects contribute to the observable signatures of accretion disks. Future work could extend this analysis by considering more complex multipole structures and accretion models, such as radiatively inefficient accretion flows, to provide further insight into the role of deformed compact objects in accretion processes. Additionally, conducting a stochastic analysis of key quantities such as radiation flux, pressure, and temperature by modelling them as Gaussian distributions can offer a probabilistic understanding of their variations. Besides,  numerical simulations and comparing theoretical predictions with observational data will further elucidate the effects of varying quadrupole moments on disk properties.





\section{Acknowledgements}

The author thanks the University of Waterloo and in part by the Government of Canada through the Department of Innovation, Science and Economic Development and by the Province of Ontario through the Ministry of Colleges and Universities at Perimeter Institute, also the research training group GRK 1620,” Models of Gravity,” funded by the German Research Foundation (DFG).

\bibliographystyle{unsrt}
\bibliography{bibqthin}
%
%

\end{document}